\documentclass[twocolumn,noarxiv,floats,floatfix,pra,aps,accepted=2019-11-15]{quantumarticle}

\pdfoutput=1

\usepackage{amsfonts}
\usepackage{amssymb}
\usepackage{amsmath}
\usepackage{calc}
\usepackage{subfigure}
\usepackage{graphicx}
\usepackage{epstopdf}
\usepackage{dcolumn}
\usepackage{bm}
\usepackage{color} 
\usepackage{CJK}
\usepackage{multirow}
\usepackage[dvipsnames]{xcolor}

\begin{document}

\title{Graded-index  optical fiber emulator of an interacting three-atom system: illumination control of particle statistics and classical non-separability}

\author{M.A.~Garc\'ia-March}
\affiliation{ICFO -- Institut de Ciencies Fotoniques, The Barcelona Institute of Science and Technology, Av.\ Carl Friedrich Gauss 3, 08860 Castelldefels (Barcelona), Spain}
\affiliation{Instituto Universitario de Matem\'atica Pura y Aplicada, Universitat Polit\`ecnica de Val\`encia, E-46022 Val\`encia, Spain}
\orcid{0000-0001-7092-838X}

\author{N.L.~Harshman}
\affiliation{Department of Physics, American University, 4000 Massachusetts Avenue NW, Washington, DC 20016, USA}
\orcid{0000-0003-2655-3327}

\author{H.~da Silva}
\affiliation{Universidade Federal de Itajub\'a, Av. BPS 1303, Itajub\'a, Minas Gerais 37500-903, Brazil}
\orcid{0000-0002-3002-7736}

\author{T.~Fogarty}
\affiliation{Quantum Systems Unit, Okinawa Institute of Science and Technology Graduate University, 1919-1 Tancha, Onna, Okinawa
904-0495, Japan}
\orcid{0000-0003-4940-5861}

\author{Th.~Busch}
\affiliation{Quantum Systems Unit, Okinawa Institute of Science and Technology Graduate University, 1919-1 Tancha, Onna, Okinawa
904-0495, Japan}
\orcid{0000-0003-0535-2833}

\author{M.~Lewenstein}
\affiliation{ICFO -- Institut de Ciencies Fotoniques, The Barcelona Institute of Science and Technology, Av. Carl Friedrich Gauss 3, 08860 Castelldefels (Barcelona), Spain}
\affiliation{ICREA, Pg. Lluis Companys 23, 08010 Barcelona, Spain}
\orcid{0000-0002-0210-7800}

\author{A.~Ferrando}
\affiliation{Department d'Optica. Universitat de Val\`encia, Dr. Moliner, 50, E-46100 Burjassot (Val\`encia), Spain}
\orcid{0000-0003-2002-4846}

\begin{abstract}
We show that a system of three trapped ultracold and strongly interacting atoms in one-dimension can be emulated using an optical fiber with a graded-index profile and thin metallic slabs. While the wave-nature of single quantum particles leads to direct and well known analogies with classical optics, for interacting many-particle systems with unrestricted statistics such analoga  are not straightforward. Here we study the symmetries present in the fiber eigenstates by using discrete group theory and show that, by spatially modulating the incident field, one can select the atomic statistics, i.e., emulate a system of three bosons, fermions or two bosons or fermions plus an additional distinguishable particle. We also show that the optical system is able to produce classical non-separability resembling that found in the analogous atomic system. 
\end{abstract}

\maketitle

\section{Introduction}

After the successful quest for preparing and measuring single quantum particles (see for example \cite{HarocheNobel,WinelandNobel}), the next task is to achieve the same kind of control over quantum systems with increasing degrees of complexity.  This will further advance our understanding of  fundamental quantum mechanics and is also predicted boost the possibilities offered by modern quantum technologies.  However, due to the exponential increase of the size of the Hilbert space for many particle systems, this is a formidable task in which theoretical and experimental progress must go hand in hand, assisting each other to facilitate scientific accomplishments.

Cold atomic systems have been at the forefront of this quest for the last two decades and by today a large effort into understanding few particle systems exists theoretically and experimentally \cite{1DReviewMAGM}. However, the experimental challenge is still very large and especially measuring  small systems reliably remains as an arduous quest. One strategy to mitigate these are to take advantage of experimental simulators, which are setups that are easier to control but follow the same dynamics as the original system. For quantum mechanical systems one can exploit the well-known similarity between the matter wave nature of particles and classical wave optics, which is based on the fact that the paraxial wave equation for monochromatic light propagating along the paraxial direction $z$ of an optical waveguide is of the same form as the Schr\"odinger equation. This similarity has been exploited in many examples in the past~\cite{nienhuis_paraxial_1993,chavez-cerda_quantumlike_2007,2008LonguiLPR}, however mostly for single particle dynamics.

\begin{figure}[tbp]
    \centering
     \includegraphics[width=0.99\columnwidth]{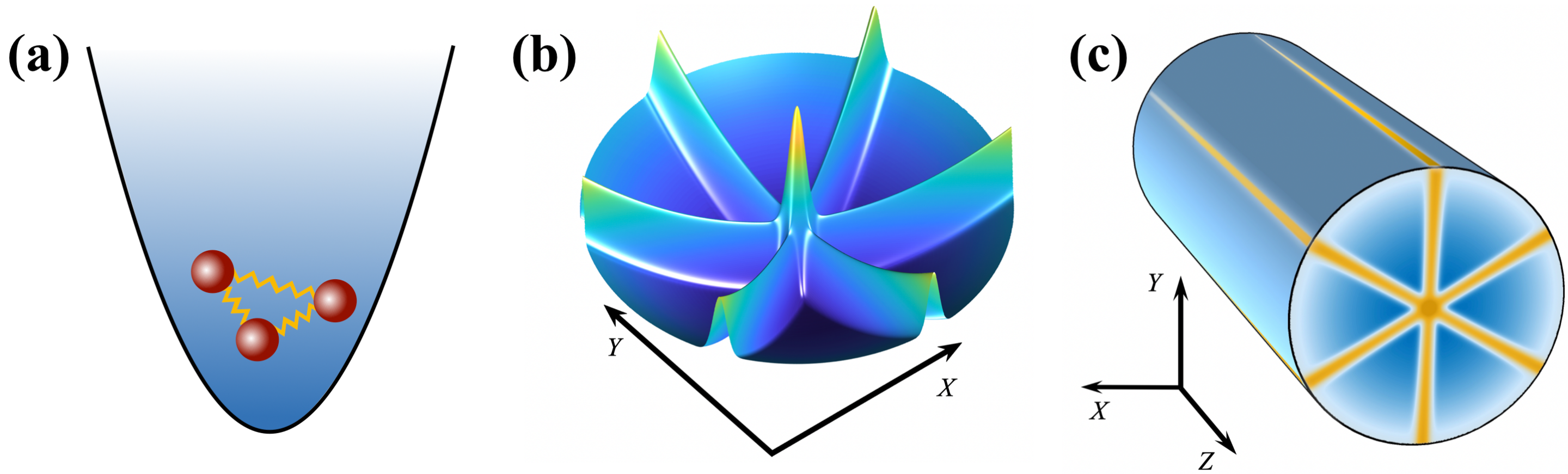}    
    \caption{Schematic of the system. (a) Three interacting atoms in a one-dimensional harmonic trap. (b) Representation of the refractive index in the $x-y$ plane. (c) Schematic of the fiber.   \label{fig:fig1} }
\end{figure}

Here we consider a one-dimensional quantum system of three harmonically trapped atoms  interacting through a strong, short-range potential and show that an analogy with a graded-index (GRIN) optical fiber with three thin slabs of a metallic material in an hexagonal geometry  exists (see Fig.~\ref{fig:fig1}). The paraxial propagation of a polarized monochromatic laser beam in such a fiber is described by a wave propagation equation which is Schr\"odinger-like and often called the Fock-Leontovich equation~\cite{fedele_beam_2003,manko_analogs_2008}. The longitudinal dimension  along the fiber plays the role of time and the inhomogeneous refractive GRIN index profile of the fiber plays the role of an external potential.  We will show below that the thin metallic slabs can play the role of the contact interactions between the atoms and that by properly designing the spatial profile of the incident laser beam  it is possible to select the statistics of the atoms emulated, that is, if they resemble bosons, fermions, or mixtures.  
  
We emphasize here that the characterization of the  modes guided by the GRIN fiber with three thin metallic slabs is of interest in itself for the optics community, independently of the analogy with the quantum system of three atoms. Graded-index fibers are multimode fibers, that is, they can propagate several modes~\cite{1973Gloge,Sodha1977,Snyder1983}. There is a recent revival in the interest in these kinds of fibers, as they have been identified as very versatile systems to study spatio-temporal non-linear effects~\cite{Longhi2003,2012Mafi}. A non-comprehensive list of recent works include the observation of optical solitons and soliton self-frequency lifting~\cite{Renninger2013}, the generation of ultrashort  pulses and even supercontinuum~\cite{Wright2015}, or beam self-cleaning~\cite{Wright2016,Krupa2017}. However, the description of pulse propagation in these fibers is rather difficult, as it must include both the three spatial dimensions and time to correctly capture the non-linear dynamics of multiple co-propagating modes  (for a simplified model see~\cite{Conforti2017}).  Yet, GRIN fibers represent an ideal set-up for a variety of phenomena  in complexity science, due to the collective dynamics associated  with the interplay between disorder, dissipation, and non-linearity~\cite{Wright2016}. Here we do not consider spatio-temporal dynamics or non-linearities, as we detail later. However, multimode GRIN fibers with thin metallic slabs allow for both to be included in future work.

As our model is an example of an analogy between a classical and a quantum system, an inferred property for the target optical system from the source quantum system is the existence of classical entanglement~\cite{1998SpreeuwFoP,spreeuw_classical_2001,2014GhoseRTS,2019KorolkovaRPP}. Classical entanglement occurs in a wide variety in optical systems, is not restricted to those described by the Fock-Leontovich equation,  and often includes polarization degrees of freedom~\cite{aiello_quantumlike_2015,2009LuisOC}. It has been proposed that a better name for this property is {\it classical  non-separability}~\cite{2015KarimiScience}, because the classical target system lacks the potential non-locality  of quantum systems with entanglement~\cite{2009LuisOC}. It is also worth stressing that classical entanglement cannot be used as a resource for applications in quantum information theory. In our system non-locality is associated with a measurement of an entangled system, which when taken in one region of space  dictates the outcome in another region. In this sense one can distiniguish two types of entanglement~\cite{1998SpreeuwFoP,spreeuw_classical_2001,harshman_tensor_2007,aiello_quantumlike_2015}: (i) intersystem entanglement (or true-multiparticle entanglement) and (ii) intrasystem entanglement (between different degrees of freedom of a single particle). It is commonly accepted that intersystem entanglement can only occur in quantum systems as it can lead to non-locality. The examples of classical non-separability found in literature are mostly associated with two different degrees of freedom of the same particle, and a remarkable realization classically non-separable states with three degrees of freedom were done using path, polarization and transverse modes~\cite{2016BalthazarOL}.  Below we show how in the system we introduce here  classical non-separability between different particles occurs in a scalar system. In this sense it is an analogous to type (i) entanglement (intersystem), but as the measurement problem remains, it does not lead to non-locality.  We note that there is a set of works where the goal is to use classical fields to reproduce non-classical correlations between different measurements~\cite{lee_experimental_2002,fu_classical_2004,2004LeePRA,2018DeZelaOptica}, including simulations of Bell-like inequalities~\cite{2015StoklasaNJP,2015QianOptica}.   

Our manuscript is structured as followed. In Section~\ref{sec:fiber} we detail the characteristics of the fiber under study. We perform a full modal analysis of it and classify the modes according to the rotational discrete symmetry of the system. The analogy with the atomic system is constructed in Section~\ref{sec:analogy} and we show how the wave function can be interpreted as giving information of the ordering of the particles. In Section~\ref{sec:classnonsep} we discuss the non-separability of the classical states and in Section~\ref{sec:conclusions} we conclude by laying out possible applications and further developments of this system. Two appendixes provide supplementary details about the symmetry methods we employ and about the Bose-Fermi mapping.

\section{Optical system: GRIN fiber with three thin metallic slabs}
\label{sec:fiber}

The paraxial propagation of a monochromatic optical beam of constant polarization along a  fiber with an inhomogeneous refractive index profile is given by
\begin{equation}
    \label{intro1}
-i2n_{0}k_{0}\frac{\partial}{\partial \tilde z}\tilde{\Phi}  =\left[\nabla_{t}^{2}+k_{0}^{2}\left(\tilde n^{2}\left(\tilde x,\tilde y\right)-n_{0}^{2}\right)\right]\tilde{\Phi},
\end{equation}
where $\tilde  z$ is the axial coordinate of the fiber, $\{\tilde x,\tilde y\}$ are the transverse coordinates, $\nabla_{t}^{2}$ is the Laplacian in the transverse coordinates, $\tilde n(\tilde x,\tilde y)$ is the index of refraction profile with a reference value of $n_{0}$, and $k_{0}$ is the wave number. To facilitate comparison with the Schr\"odinger equation, we remove the length units by dividing  by $-2n_{0}k_{0}^{2}$ 
\begin{equation}
\label{intro2}
i\frac{\partial}{\partial z} \Phi=\left(-\frac{1}{2}\nabla_{x,y}^{2}+\Delta n\left(x,y\right)\right) \Phi,
\end{equation}
where, $\{ x, y, z\}$ are the dimensionless coordinates $x  =k_{0}\sqrt{n_{0}} \tilde x$, $y =k_{0}\sqrt{n_{0}} \tilde y$, $z  =k_{0} \tilde z$, $\Phi(x,y,z) = \tilde\Phi(\tilde{x},\tilde{y},\tilde{z})$, $n(x,y) = \tilde{n}(\tilde{x},\tilde{y})$ and 
\begin{equation}
\label{eq:delta_n2}
\Delta n\left(x,y\right)=[n_{0}^{2} - n^{2}\left(x,y\right)]/2n_{0}.
\end{equation}
When the refractive index profile remains close to the reference index $n_0$, eq.~(\ref{eq:delta_n2}) simplifies to $\Delta n\left(x,y\right) \approx n_{0} -  n\left(x,y\right)$.

The form of eq.~\eqref{intro2} mimics the two-dimensional time-dependent Schr\"odinger equation with $z$ playing the role of time and $\Delta n\left(x,y\right)$ the role of a potential energy. Making the substitution $\Phi(x,y,z) = \exp(-i \mu z) \phi(x,y)$ to separate the longitudinal and transverse coordinates, one can see that solving for the transverse optical modes $\phi(x,y)$ and paraxial propagation constant $\mu$ is equivalent to solving for the energy spectrum of a quantum Hamiltonian with two degrees of freedom
\begin{equation}\label{eq:optschr}
\hat{H} \phi(x,y) \equiv \left(-\frac{1}{2}\nabla_{x,y}^{2}+\Delta n\left(x,y\right)\right)\phi(x,y) =  \mu \phi(x,y).
\end{equation}
This analogy between fiber optics in the paraxial approximation and two-dimensional quantum mechanics is well-known (see e.g. \cite{krivoshlykov_optical_1980}) and eq.~\eqref{intro2} is indeed sometimes called the optical Schr\"odinger equation~\cite{joseph_effect_2015}. Here, we  assume a longitudinally homogeneous fiber; relaxing this requirement allows the simulation of quantum systems with time-varying mass or potential.

\subsection{GRIN fiber optical modes}

We build our effective potential for the analogy by combining GRIN fibers with metallic sectioning. A GRIN fiber has a refractive index $n\left(x,y\right)$ that decreases continuously with the radial distance to the optical axis of the fiber. Here we consider the particular case of a parabolic profile that focuses the beam and provides guidance in the fiber (also called selfoc fibers~\cite{manko_analogs_2008}), i.e.
\begin{equation}
 \label{eq:refractiveindexG}
 \Delta n_{\rm{GRIN}}\left(x,y\right)\!= \left\{\begin{array}{cc}
                                                                 \frac{1}{2}\!\left(x^2+y^2\right) & \rho <R\\
                                                                 \frac{1}{2}R^2 & \rho \ge R
                                                                \end{array}\right., 
\end{equation}
where $\rho=\sqrt{x^2+y^2}$. These kinds of GRIN fibers have previously been proposed to simulate two-dimensional quantum oscillators~\cite{nienhuis_paraxial_1993,manko_quantum_2001}.  

For a fiber with transverse index $\Delta n_{\rm{GRIN}}$, eq.~(\ref{intro2}) is separable into radial $\rho = \sqrt{x^2+y^2}$ and polar $\theta =\arctan(y/x)$ coordinates. The boundary at $\rho = R$ can be ignored for the lowest modes and in this approximation the optical Schr\"odinger equation (\ref{eq:optschr}) describes a two-dimensional isotropic harmonic oscillator. Separating in polar coordinates, the corresponding solutions for the mode functions in polar coordinates $|\ell,\nu\rangle$ are given by 
\begin{equation}
\label{eq:l,nrho}
\phi_{\ell,\nu}(\rho,\theta)=N \rho^{|\ell|} L^{|\ell|}_{\nu}(\rho^2) e^{-\rho^2/2}e^{i\ell\theta},
\end{equation}
with $L^{|\ell|}_{\nu}(z)$ the generalized Laguerre polynomial and normalization constant $N=\sqrt{\nu!/\pi(\nu+|\ell|)!}$. These modes (\ref{eq:l,nrho}) correspond to Laguerre-Gaussian modes centered at the origin and the mode indices correspond to orbital angular momentum (OAM) $\ell=0,\pm 1,\pm 2,\cdots$ and the number of radial nodes $\nu=0,1,\cdots$. The OAM $\ell$  gives the charge of the central singularity and is the quantum number for the $\mathrm{O}(2)$ symmetry of the isotropic oscillator. The propagation constant (analogous to energy) of the mode $|\ell,\nu\rangle$ is $\mu = 2 \nu +|\ell| +1 $ and except for the lowest mode $|\ell,\nu\rangle = |0,0 \rangle$, all modes are degenerate with degeneracy $d(\mu) = \mu$. 

\subsection{GRIN fiber and metallic slabs}

Next, we section the fiber longitudinally with thin slabs of metal. For later applications to three-particle systems, we consider the case of three slabs that split the fiber into six identical sectors (see Fig.~\ref{fig:fig1}). This is described by adding to $\Delta n_{\rm{GRIN}}$ an additional term formed by three Gaussians  of width $\sigma$
\begin{align}
 \label{eq:refractiveindexC6}
 \Delta n_{\mathcal{C}_{6v}}\left(x,y\right)\!=\!\frac{g}{\sigma\sqrt{2\pi}}\!\Bigg\{&\exp\!\left[-\frac{x^2}{\sigma^2}\right]
    +\exp\!\left[-\frac{(x + \sqrt{3}y)^2}{4\sigma^2}\right]\nonumber\\
    &\left.+\exp\!\left[-\frac{(x - \sqrt{3}y)^2}{4\sigma^2}\right]\right\}.
\end{align}
The function $\Delta n_{\mathcal{C}_{6v}}\left(x,y\right)$  has three maxima at the lines $x=0$ and $x=\pm\sqrt{3}y$, or equivalently at $\theta \in \{ \pi/6, \pi/2, 5\pi/6, 7\pi/6, 3\pi/2, 11\pi/6\}$.

The exact form of $g$ in (\ref{eq:refractiveindexC6}) in terms of the optical parameters is crucial in the particle analogy in Section~\ref{sec:analogy} and after restoring the spatial dimensions for $\tilde{\sigma}=\sigma\,\lambda$,  one obtains
\begin{equation}
\label{eq:analogyparameters}
g=\frac{\Delta n^{\rm{max}}_{C_{6v}}(\lambda) \,\sqrt{2\pi}\tilde{\sigma}}{\lambda}.  
\end{equation}
In this work we are mainly interested in the limit where $g$ is large. Because the right hand side of eq.~\eqref{eq:analogyparameters} includes only dimensional optical parameters, the large-$g$ limit can be reached experimentally  with thin slabs of a metallic material of width $\tilde{\sigma}$. For a perfect conductor $n^2_{\rm{metal}}\to-\infty$ and therefore $ \Delta n^{\rm{max}}_{C_{6v}}=(n_0^2 - n^2_{\rm{metal}})/2n_0$ and consequently $g$ tends to infinity. One has to be careful though, as in the experimentally relevant case with a realistic metal, the dielectric constants also have an imaginary part, i.e., $\varepsilon=\varepsilon_1+ i\varepsilon_2$ and, for example, for gold at $\lambda=1500$nm, one has $\varepsilon_1 = -106.94$ and $\varepsilon_2  = 10.231 $. However, for the limit in which $g$ is large and $\tilde{\sigma}$ small the losses due to the imaginary part are small because in the regions with large $g$ the optical modes have suppressed intensity, as we show later. 

Combining the thin slabs of a metal with the GRIN fiber, the total refractive index is $\Delta n_{\rm{tot}} = \Delta n_{\rm{GRIN}} + \Delta n_{C_{6v}}$. The fiber then has six identical symmetric domains $\Omega_j$ ($j\in\{1,\ldots,6\}$) where $\Delta n_{\rm{GRIN}}$ dominates separated by the thin barriers where $\Delta n_{C_{6v}}$ dominates. This fiber profile has the six-fold symmetry of a snow flake denoted as $C_{6v}$ Sch\"onflies notation~\cite{hamermesh_group_1989}; see Appendix A for a summary of the group $C_{6v}$ and its representations~\footnote{The symmetry $C_{6v}$ is also denoted as the $D_6$ or $I_2(6)$ depending on context or application. More generally, if there are $s$ bisecting metal slabs inserted evenly, then the system has $C_{2sv}$ (aka $D_s \sim I_2(2s)$) symmetry.}.

\begin{figure}[t]
    \centering
     \includegraphics[width=0.95\columnwidth]{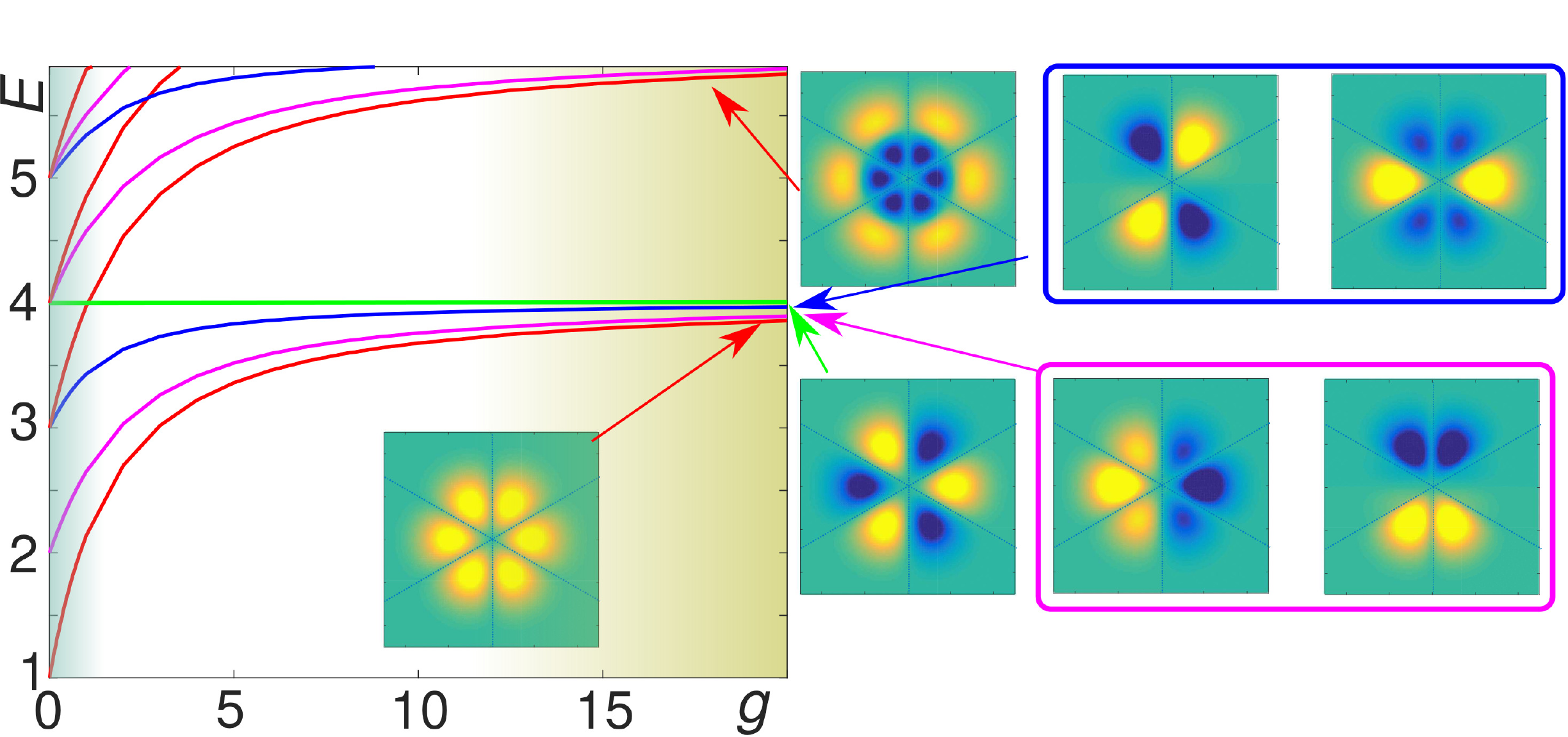}    
    \caption{Eigenenergies (propagation constant) for the optical Schr\"odinger equation (\ref{eq:optschr}) with $\Delta n_{\rm{tot}} = \Delta n_{\rm{GRIN}} + \Delta n_{C_{6v}}$ plotted against varying barrier strength $g$ for a fixed, narrow width $\sigma$. The shaded golden area highlights the region of large values of $g$, where the thin metallic slabs can be implemented with e.g. gold, as discussed in main text. The shaded green area highlights the region of low $g$, which has a weak $\Delta n_{\mathcal{C}_{6v}}$ refractive index that could be implemented with dielectric materials.  Inset and the right of the graph depicts the eigenmodes corresponding to the lowest seven modes (see Figs.~\ref{fig:fig3} and \ref{fig:fig4}). In all representations of the eigenmodes, the dotted lines indicate the position of the metallic slabs. \label{fig:fig2} }
\end{figure}

Since the barriers break the $\mathrm{O}(2)$ rotational symmetry, the angular variation of the wave function is no longer uniform and as a result orbital angular momentum $\ell$ is no longer a good modal index. Additionally, for arbitrary strengths and widths, the barriers break polar separability so $\nu$ is also not a good quantum number. However, the discrete $C_{6v}$ symmetry provides the possibility for alternate modal numbers~\cite{mcisaac_symmetry-induced_1975,guobin_mode_2003}. One useful index is called orbital angular pseudo-momentum (OAPM) and was introduced in the context of vortex solitons~\cite{ferrando_discrete-symmetry_2005,garcia-march_angular_2009}. OAPM is a discrete index $m \in \{ 0, \pm 1, \pm 2, 3\}$ that identifies how the state transforms under a discrete rotation by $\pi/3$ and it gives the charge of the central singularity~\cite{garcia-march_symmetry_2009}. In the subspace of solutions with OAPM $m$, a counterclockwise rotation by $\pi/3$ changes the phase of the optical mode $\phi(\rho,\theta)$ by $\exp(i m \pi/3)$. In the case $m=0$ the mode is symmetric with respect to $C_{6v}$ and there is no phase change from sector to sector, and when $m=3$ the mode is antisymmetric with respect to $C_{6v}$ and the phase flips from sector to sector.

Previewing the analogy with the one-dimensional, three-body system developed in the next section, the OAPM correlates to the particle content. The modes indexed by $m=0$ and $m=3$ correspond to three indistinguishable bosons or fermions. Another mode index, the reflection parity $r = \pm 1$ under reflection across the thin slab oriented along $x=0$ or $\theta = \pm \pi/2$, indicates whether particles with OAPM $m=0$ or $m=3$ are bosons or fermions. The cases of OAPM $m=\pm 1$ and $m= \pm 2$ contain solutions that describe identical but \emph{partially} distinguishable particles, such as two spin-up fermions and one spin-down fermion.

\subsection{Infinite delta-barrier limit}

\begin{figure}[t]
    \centering
     \includegraphics[width=0.8\columnwidth]{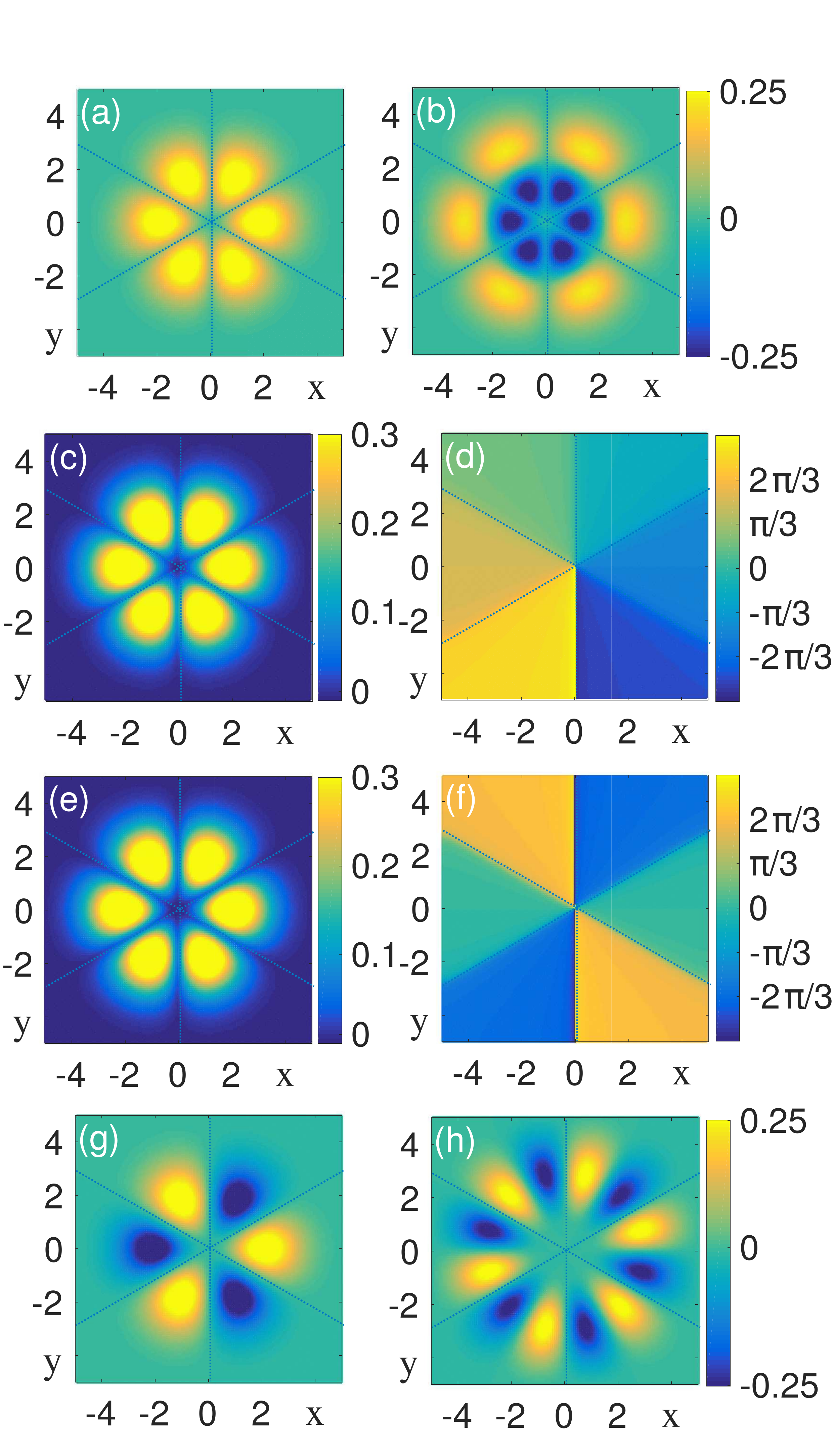}    
    \caption{Numerical eigenfunction solutions for  $g = 10$, $\sigma=0.05$ and $R\to\infty$; compare to infinite delta-barrier solutions $|m,\tilde{\nu},\tilde{\ell}\rangle$ in (\ref{eq:m,nrho,ntheta}).  (a) Ground state, $|m,\tilde{\nu},\tilde{\ell}\rangle=|0,0,0\rangle$. (b) seventh state,  $|m,\tilde{\nu},\tilde{\ell}\rangle=|0,1,0\rangle$, which carries the first radial excitation of the ground state; (c) and (d) amplitude and phase of the vortex state $|1,0,0\rangle$; (e) and (f) same for vortex state  $|2,0,0\rangle$.  (g) sixth excited state, $|m,\tilde{\nu},\tilde{\ell}\rangle=|3,0,0\rangle$;  and (h) eighteenth state, $|m,\tilde{\nu},\tilde{\ell}\rangle=|0,0,1\rangle$. \label{fig:fig3} }
\end{figure}

\begin{figure}[t]
    \centering
     \includegraphics[width=0.85\columnwidth]{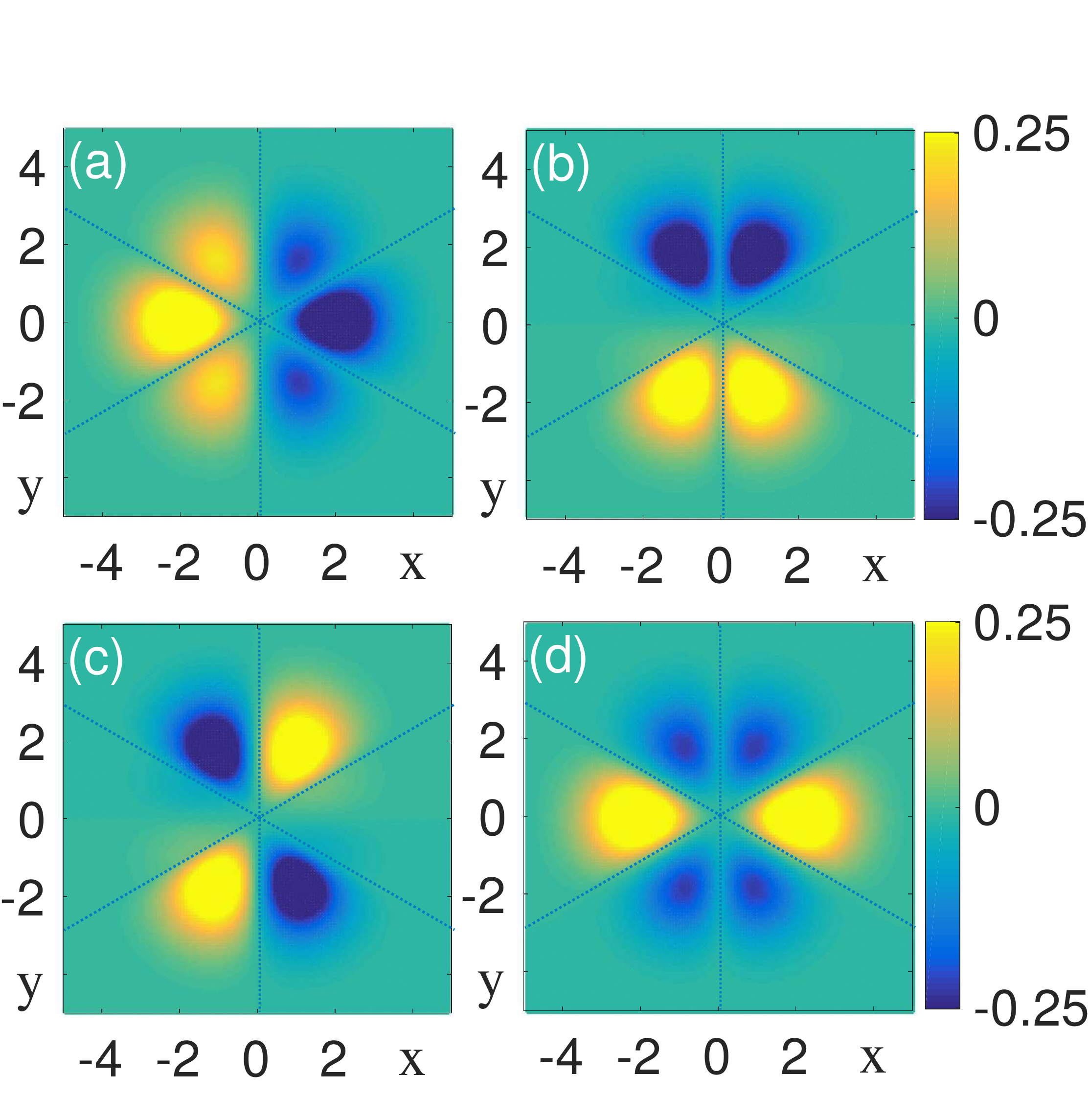}  
        \caption{Eigenfunctions for  $g = 10$ and $\sigma=0.05$.   (a) and (b), first excited doublet obtained combining the vortices with $m=1$ and $m=-1$, that is, $ |1,0,0\rangle\pm i |-1,0,0\rangle$, respectively.    (c) and (d), second excited doublet obtained as  $ |2,0,0\rangle\pm i |-2,0,0\rangle$, respectively. \label{fig:fig4}     }
\end{figure}

In the limit of infinitely narrow slabs $\tilde{\sigma} \to 0$, the Gaussian profiles in (\ref{eq:refractiveindexC6}) tend to delta functions and can be approximated as
\begin{align}
\label{eq:refractiveindexC6-delta}
\Delta n_{C_{6v}}\!&\approx\! g\left[\delta(x) + \sqrt{2}\delta(x + \sqrt{3}y) + \sqrt{2}\delta(x - \sqrt{3}y)\right]\nonumber\\
&=\frac{g}{\rho}\sum_{j=1}^6\delta\left(\theta-\frac{2j-1}{6}\right).
\end{align}
Note that the apparent singularity at $\rho=0$ is not strong enough to disrupt the self-adjointness of the effective Hamiltonian and there is no danger of ``falling to the center''~\cite{landau_chapter_1977}.
However, the potential in (\ref{eq:refractiveindexC6-delta}) does not have the correct form for polar separability for arbitrary $g$; only in the limit of impenetrable barriers $g \to \infty$ does polar separability return and we can provide exact solutions.

In the narrow, impenetrable barrier limit, each identical sector $\Omega_j$ is dynamically-isolated from the rest and within each sector approximate polar separability returns. This means the number of radial nodes by $\tilde{\nu}$ and the number of azimuthal nodes $\tilde{\ell}$ \emph{within each sector} are good mode labels (or quantum numbers). Choosing $\Omega_1 = [-\pi/6 , \pi/6]$ as the first sector, the optical mode solutions are 
\begin{equation}\label{eq:idealsector}
\phi^{1}_{\tilde{\nu},\tilde{\ell}}(\rho,\theta)=N\rho^{\tilde{\ell}}L^{\tilde{\ell}}_{\tilde{\nu}}(\rho^2) e^{-\rho^2/2}\sin[3(\tilde{\ell}+1)(\theta+\pi/6)].
\end{equation}
This equation satisfies the optical Schr\"odinger equation for a GRIN fiber (i.e., it is a special case of (\ref{eq:l,nrho})) and also satisfies the nodal boundary condition at the sectioning metal slabs when $\tilde{\nu}$ and $\tilde{\ell}$ are non-negative integers. By analogy with (\ref{eq:l,nrho}) or by direct calculation, one shows the propagation constant of this mode is $\mu = 2 \tilde{\nu} + 3\tilde{\ell} + 4$.

The mode solutions in the entire fiber can be built by using the OAPM $m$ to patch together single sector solutions like (\ref{eq:idealsector}) with the correct phase differences. An explicit expression for the mode $|m, \tilde{\nu},\tilde{\ell}\rangle$ built from sectors with $\tilde{\nu}$ radial nodes and $\tilde{\ell}$ azimuthal modes takes the form
\begin{equation}
 \label{eq:m,nrho,ntheta}
  \phi_{m,\tilde{\nu},\tilde{\ell}}(\rho,\theta)=\phi^{1}_{\tilde{\nu},\tilde{\ell}}\left(\rho,\theta - \theta_j\right) e^{-im \theta_j}\ \mbox{for}\ \theta\in\Omega_j
 \end{equation}
where $\theta_j = (j-1)\pi/3$  and  $\Omega_j = [(2j-3)\pi/6 , (2j-1)\pi/6]$. The six ways to choose the relative phases and paste the section functions together such that the state respects  $C_{6v}$ symmetry are precisely the six possible values the OAPM $m$ takes: $m=0,\pm1,\pm2$ and $3$. The six states  $|m, \tilde{\nu},\tilde{\ell}\rangle$ with the same $\tilde{\nu}$ and $\tilde{\ell}$ are degenerate and have the same propagation constant $\mu = 2 \tilde{\nu} + 3\tilde{\ell} + 4$ independent of OAPM $m$. In this limit, the effective Hamiltonian is superintegrable, i.e.~there are more integrals of motion than degrees of freedom~\cite{andersen_hybrid_2017}. This degeneracy is only present in the idealized case of delta-barriers and infinite $g$. For both the idealized finite-$g$ delta-barrier potential (\ref{eq:refractiveindexC6-delta}) and the more realistic Gaussian potential (6), the tunneling among sectors lifts the degeneracy so that states with different $|m|$ have different propagation constants.

To show how this works, we calculated numerically the eigenfunctions  in the presence of the metal slabs of height $g=10$, width $\sigma=0.05$, and $R$ larger than the size of the computational domain (a box of side $L=10$). As shown in Fig.~\ref{fig:fig2}, in this limit the six modes with different $m$ and same $\tilde{\nu}$ and $\tilde{\ell}$ are quasi-degenerate and approximate the infinite delta-barrier solutions (\ref{eq:m,nrho,ntheta}). A selection of modes are depicted in Fig.~\ref{fig:fig3}, including \ref{fig:fig3}(a) the ground state mode $|0,0,0\rangle$; \ref{fig:fig3}(b) the lowest energy state mode with one radial excitation $|0,1,0\rangle$; and \ref{fig:fig3}(h) the highest energy state mode with one polar excitation $|0,0,1\rangle$.

Subfigures \ref{fig:fig3}(c)-\ref{fig:fig3}(g) depict three other modes with $\tilde{\nu}=0$ and $\tilde{\ell}=0$. The $C_{6v}$ symmetry ensures that pairs of modes with $|m|=1$ and with $|m|=2$ are degenerate, so we only depict the magnitude and phase of the $m=1$ in \ref{fig:fig3}(c)-\ref{fig:fig3}(d) and  the magnitude and phase of $m=2$ in \ref{fig:fig3}(e)-\ref{fig:fig3}(f).
Because these modes are degenerate, instead of working with the complex modes $|\pm 1,\tilde{\nu},\tilde{\ell}\rangle$ and $|\pm 2,\tilde{\nu},\tilde{\ell}\rangle$ we can take take linear combinations such as $|1,\tilde{\nu},\tilde{\ell}\rangle \pm i |-1,\tilde{\nu},\tilde{\ell}\rangle$ that result in real modes.  Examples are presented in Fig.~\ref{fig:fig4} that show that these real modes are no longer OAPM eigenstates of $\pi/3$ rotations, but they diagonalize into a pair of orthogonal reflections and have $C_{2v}$ symmetry.

\section{Optical analogy to the three-particle model}
\label{sec:analogy}

In this section, we show how the fiber introduced above can be used to simulate a specific quantum system of current interest in ultracold atomic physics: three interacting atoms trapped in a one-dimensional harmonic potential with strong, short-range interactions (see e.g. the striking experiments in~\cite{2011Serwane,2013Wenz}). We also show that the optical modes of the fiber can simulate the wave functions of energy eigenstates for any particle statistics, including single-species and multi-species fermions and bosons. For this the OAPM modal number $m$ and reflection parity $r$ play the role of effective statistical parameter. 

\subsection{The three particle Hamiltonian}

To see that the optical Schr\"odinger equation for the fiber above can simulate a three-body, one-dimensional system, let us start by considering the quantum Hamiltonian for three interacting particles in a one dimensional harmonic trap
\begin{equation}
    \label{model31}
    H = \frac{\hbar \omega}{2} \sum_{i=1}^3 \left( -\frac{d^2}{dx_i ^2}+ x_i^2 \right) + \sum_{i<j} V_{ij}\left(|x_i - x_j|\right).
\end{equation}
For convenience, we have scaled all distances by the harmonic oscillator length $a = \sqrt{\hbar/(m \omega)}$ and the coordinates $x_i$ are the unitless positions of the three particles. 
The two-body interaction depends only on the distance between pairs of particles.
Next we perform a transformation from the particle positions coordinates to the normalized Jacobi coordinates
\begin{align}\label{jacobi}
    R&=\frac{x_1+ x_2 + x_3}{3},\\
     x&=\frac{x_1-x_2}{\sqrt{2}},\\
     y&=\frac{x_1+x_2}{\sqrt{6}}-\sqrt{\frac{2}{3}}x_3,
\end{align}
where $R$ is proportional to the center-of-mass and $x$ and $y$ are a specific but arbitrary choice for the orientation of three-body relative coordinates.
With this,  the Hamiltonian in eq.~\eqref{model31} can be split into a center-of-mass and a relative part, $H=H_\mathrm{cm}+H_\mathrm{rel}$, with
\begin{subequations}
\begin{align}  
    H_\mathrm{cm}=&-\frac{\hbar\omega}{2}\frac{d^2}{dR^2}+\frac{\hbar\omega}{2} R^2\,,\label{model34_COM}\\
    H_\mathrm{rel}=&-\frac{\hbar\omega}{2}\left(\frac{d^2}{dx^2}+\frac{d^2}{dy^2}\right)+\frac{\hbar\omega}{2} (x^2+y^2) + V_\mathrm{int}(x,y)\,,\label{model34_rel} 
\end{align}
and
\begin{align}
    V_\mathrm{int}(x,y)= V_{12}\left(\sqrt{2} |x|\right) + &V_{13}\left(\!\frac{|x \!+\! \sqrt{3}y|}{\sqrt{2}}\!\right) \nonumber\\
         &+ V_{23}\left(\!\frac{|x \!-\! \sqrt{3}y|}{\sqrt{2}}\!\right).
\end{align}
     \end{subequations}
This transformation therefore separates out the trivial center-of-mass degree of freedom whose dynamics are described by the one-dimensional oscillator $H_\mathrm{cm}$. The remaining two relative degrees of freedom are described by $H_\mathrm{rel}$ that has the same six-fold symmetry of the previous section. 

If we now take $V_{ij}$ to be given by a narrow Gaussian
\begin{equation}
   V_{ij}(z) = \frac{g}{\sigma\sqrt{2\pi}} \exp\!\left[-\frac{z^2}{2\sigma^2}\right],
\end{equation}
we recover the effective Hamiltonian given by the fiber mode propagation equation of the previous section with  $\Delta n_{\rm{tot}} = \Delta n_{\rm{GRIN}} + \Delta n_{C_{6v}}$ and all the analysis of the previous section holds. In the limit of highly localized and strong scattering potentials, the modes $|m,\tilde{\nu},\tilde{\ell}\rangle$ become exact and all six states with the same $\tilde{\nu}$ and $\tilde{\ell}$ become six-fold degenerate again.

\subsection{Particle permutation symmetry, OAPM and particle statistics}

\begin{figure}[t]
    \centering
     \includegraphics[width=0.8\columnwidth]{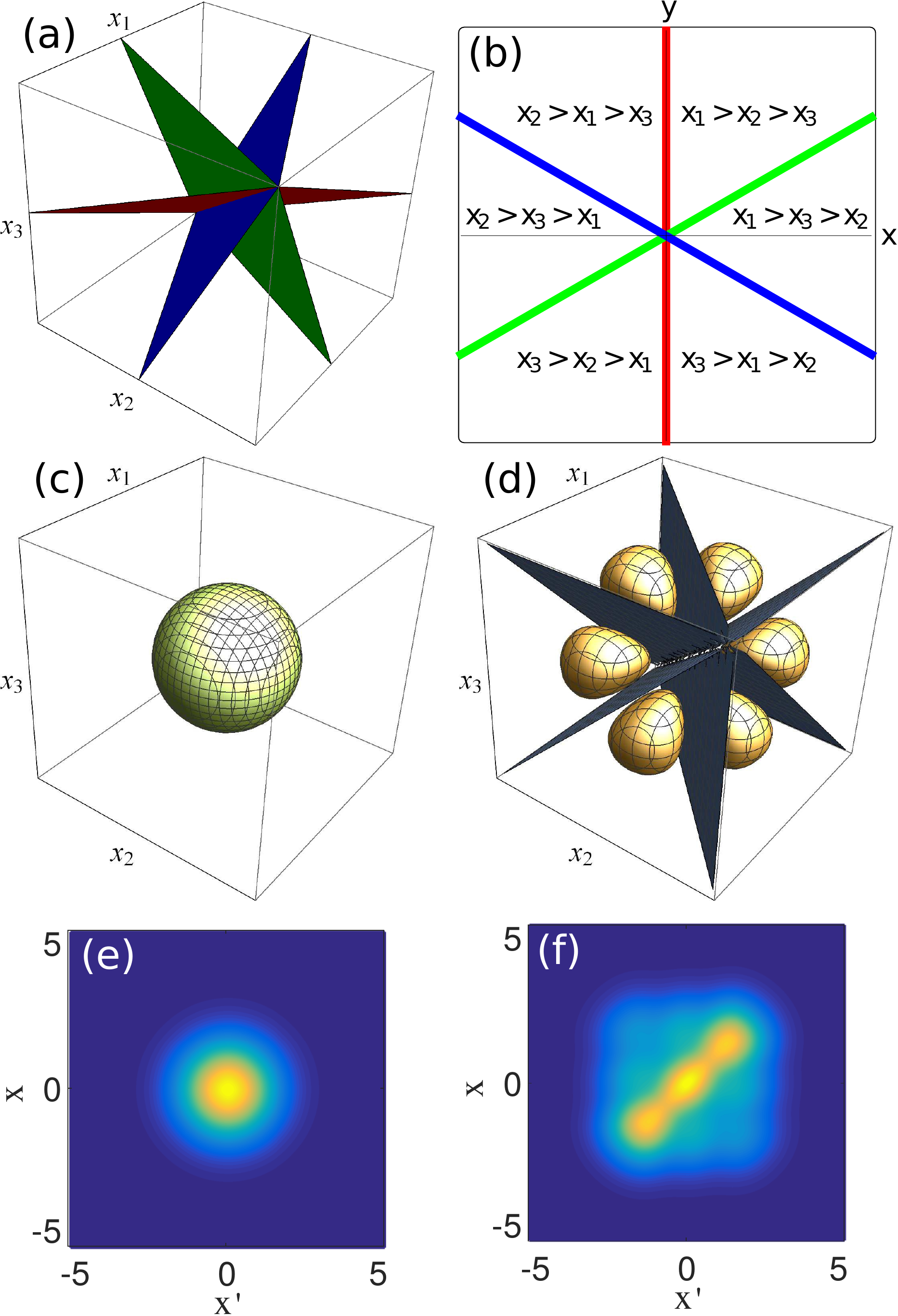}    
    \caption{(a) The full $(x_1,x_2,x_3)$ three-particle configuration space. The three planes represent the two-body coincidences $x_1=x_2$ (red), $x_2=x_3$ (green) and $x_3=x_1$ (blue). (b) The relative $(x,y)$ configuration space defined by the orthogonal Jacobi transformation~(\ref{jacobi}). The lines are the projection of the planes in subfigure (a). Each of the six sectors corresponds to specific orderings of three particles. Reflecting across the two-particle coincidence lines is equivalent to a pairwise exchange of identical particles. Complete BBB wave function $\Psi(x_1,x_2,x_3,z)$ for the non-interacting (c) and impenetrable delta-function barrier limits (d). Sub figure (e) and (f) are the corresponding OBDM in each limit, respectively.  \label{fig:Fig5} }
\end{figure}

Like the metallic slabs section the fiber into six sectors, the two-body interactions section the $(x,y)$ relative configuration space of the three particles into six sectors. Each section corresponds to the particles being in a specific order (see Fig.~\ref{fig:Fig5}). In a model with distinguishable particles, the phase relation between different orderings of particles is unconstrained. In contrast, three identical bosons must be symmetric under a particle exchange and three identical fermions must be antisymmetric. As a result, sectors corresponding to different orders must have specific phase relations if they are to represent the solutions of identical particles. Conveniently, the OAPM $m$ and reflection parity $r$ that derive from the $C_{6v}$ symmetry can be used as parameters that account for particle statistics~\cite{garcia-march_distinguishability_2014}.

In the original Hamiltonian (\ref{model31}), the particle permutation symmetry is evident: one can permute the coordinates $(x_1,x_2,x_3)$ without changing the form of the Hamiltonian. The group of particle permutations is called $S_3$ and we denote the operators that represent these transformations by $\hat{\sigma}_p$ for $p\in S_3$. For example, the operator $\hat{\sigma}_{213}$ exchanges particles 1 and 2, the operator $\hat{\sigma}_{312}$ cycles $(x_1,x_2,x_3)$ into $(x_3,x_1,x_2)$, and the operator $\hat{\sigma}_{123}=\hat{e}$ is the identity. Additionally, the parity inversion $(x_1,x_2,x_3) \rightarrow (-x_1,-x_2,-x_3)$ leaves the Hamiltonian invariant. We denote the parity inversion operator by $\hat{\pi}$ and the two-element group it generates by $Z_2$. Because parity inversion and particle permutations commute, the total symmetry group is the direct product $S_3 \times Z_2$; see Appendix A for an enumeration of all twelve elements of this symmetry group. 

When restricted to the relative plane, the particle permutations and parity are realized as the transformations in $C_{6v}$. For example, the pairwise exchange $\hat{\sigma}_{213}$ is the reflection across $x=0$ that maps $(x,y)$ into $(-x,y)$. The other two pairwise exchanges $\hat{\sigma}_{321}$ and $\hat{\sigma}_{132}$ are also realized as reflections in the relative $(x,y)$-plane along the lines $x=-\sqrt{3}y$ and $x=\sqrt{3}y$, respectively. The two three-cycles $\hat{\sigma}_{312}$ and $\hat{\sigma}_{231}$ are rotations by $+2 \pi/3$ and $-2 \pi/3$, respectively, and parity $\hat{\pi}$ is a rotation by $\pi$. Finally, combining parity $\hat{\pi}$ and the three-cycle $\hat{\sigma}_{231}$, the symmetry transformation $\hat{c}_6 = \hat{\pi}\hat{\sigma}_{231}$ is a rotation by $+\pi/3$. 

Therefore by looking at how optical modes transform under the reflections and rotations in $C_{6v}$, we also analyze how the analogous three-particle wave function transforms under particle permutations and parity $S_3 \times Z_2$. In fact, the mode numbers OAPM $m\in\{0,\pm 1, \pm 2, 3\}$ and reflection parity $r$ introduced in the previous section are the eigenvalues of the operators  $\hat{c}_6$ and $\hat{\sigma}_{213}$, respectively. Using them we build a classification of particle statistics as follows:
\begin{itemize}
\item The energy subspaces with $(m,r)=(0,1)$ or $(3,1)$ are non-degenerate modes with the requisite symmetry to realize states of three indistinguishable bosons, denoted BBB modes. These states also could represent identical but distinguishable particles.
\item The energy subspaces with $(m,r)=(0,-1)$ or $(3,-1)$ are also non-degenerate. These wave functions have the requisite symmetry to be states of three indistinguishable fermions, denoted FFF modes. As before, these states also could represent identical but distinguishable particles.
\item The energy subspaces labelled by $|m|= 1$ or $|m|= 2$ correspond to doubly-degenerate modes. In these two-dimensional subspaces, the operators $\hat{c}_6$ and $\hat{\sigma}_{213}$ do not commute and the OAPM $m$ and reflection parity $r$ cannot be simultaneously diagonalized. Choosing to diagonalize $\hat{\sigma}_{213}$ within the $|m|= 1$ or $|m|= 2$ subspace, there are superpositions that describe states with $r=1$ that are symmetric under exchanges of particles 1 and 2, but have no fixed phase relation for a pairwise exchange including particle 3. We call this a BBX mode, because it can describe the state of two identical bosons and one distinguishable third particle. Similarly, there are FFX modes where the exchange of two identical fermions is antisymmetric with $r=-1$  and the third particle is distinguishable. 
\end{itemize}
Additionally, from the relation between parity inversion and six-fold rotation $\hat{\pi} = (\hat{c}_6)^3$, states with OAPM $m$ have parity inversion $\exp(i m \pi/3)^3=(-1)^m$. More details on how to derive these results are provided in~\cite{harshman_symmetries_2012, harshman_one-dimensional_2016, andersen_hybrid_2017} and Appendix A.

In light of this assignment of OAPM and reflection parity to possible combinations of identical particles, the mode level structure depicted in Fig.~\ref{fig:fig2} reveals further insights. First, note because the phase of FFF modes must change sign at the section boundaries, the density must drop to zero and, as a result, FFF modes do not `feel' the interaction strongly (or at all, in the impenetrable delta-function barrier limit). The energies of these FFF modes are therefore nearly horizontal even as the interaction strength $g$ is increased. In contrast, the symmetric BBB modes experience the greatest variation with $g$, and in the large $g$ converge to the same energy as an FFF mode with the same wave function in the sector (i.e.~same $\tilde{\nu}$ and $\tilde{\ell}$).  This is reminiscent of the famous Bose-Fermi mapping for infinite strength contact interactions, first identified by Girardeau~\cite{girardeau_relationship_1960}; see Appendix B for more details.

Proper illumination of the fiber then permits to select the appropriate mode.
For instance, to excite a BBB mode one can illuminate with a structured beam. For example, one can illuminate with an intensity modulation that follows the  $C_{6v}$ symmetry of the BBB mode with $m=0$, $\tilde{n}=0$ and $\tilde{\ell}=0$. To excite the FFF mode with $m=3$, $\tilde{n}=0$ and $\tilde{\ell}=0$  one has to modulate not only the intensity but also to imprint a phase jump of $\pi$ between sectors, which can be achieved by using spatial phase modulators. For BBX or FFX modes with $|m|=1$ or $2$, the input beam  has to mimic  the $C_{2v}$ symmetry and the $\pi$ phase jumps as in Fig.~\ref{fig:fig4}.

\section{Interpretation of the optical mode amplitude as a many-body wave function: classical non-separability}
\label{sec:classnonsep}

In this section we discuss how to extract the information on classical non-separability from the optical field $\Phi(x,y,z)$ at a certain distance $z$. To reconstruct the function in the
 $(x_1,x_2,x_3)$ configuration space, one must account for the center of mass and its  evolution along $z$. This is given by the modes of the one-dimensional  harmonic oscillator, which we label with the quantum number $n_{\rm{R}}$,  and denote as
  $\varphi_{n_{\rm{R}}}(R)$, so that the total wave function is $\Psi(x,y,R,z)=\Phi(x,y,z)\varphi_{n_{\rm{R}}}(R)\exp[-i n_{\rm{R}}z]$.  From $\Psi(x,y,R,z)$, one can then transform back to the variables $(x_1,x_2,x_3)$ to get $\Psi(x_1,x_2,x_3,z)$, which can be performed digitally after phase and amplitude detection of $\Phi(x,y,z)$.
	
With the total wave function $\Psi(x_1,x_2,x_3,z)$, the classical  non-separability can be evaluated by first calculating the one body density matrix (OBDM), defined as
 \begin{equation}
 \rho(x,x')=\int \Psi^*(x,x_2,x_3) \Psi (x',x_2,x_3)\,d x_2\,d x_3,
 \end{equation}
 and normalized to one. 
 The classical non-separability is then defined by the von Neumann entropy 
\begin{equation}
 S_{\rho(x,x')}=-\mbox{Tr}[\rho(x,x')\ln\rho(x,x')]=-\sum_j \lambda_j \ln\lambda_j,
\end{equation}
where we have denoted the eigenvalues of $\rho(x,x')$ as $\lambda_j$. We recall that the von Neumann entropy is zero for a pure state (non-mixed) and maximal and equal to $\rm{ln}(3)$ for a maximally mixed state (maximal non-separability). 

Let us illustrate the interpretation of the classical non-separability for a system of three bosons. In this case the ground state wave function for the non-interacting system is $ \Psi _{\rm{B,gs}}^{g=0}(x_1,x_2,x_3)=C\left[\prod_{i=1}^3 e^{-x_i^2/2} \right]$, with $C$ being a normalization constant (see Fig.~\ref{fig:Fig5} (c)). In the impenetrable delta-function barrier limit, the wave function is
\begin{equation}
\label{eq:Jastrow3}
 \Psi_{\rm{B,gs}}^{g=\infty}(x_1,x_2,x_3)=C\left[\prod_{i=1}^3 e^{-x_i^2/2} \right]\prod_{1\le j < k \le 3}|x_k-x_j|,
\end{equation}
which is the solution obtained from the Bose-Fermi mapping theorem (see Fig.~\ref{fig:Fig5} (d) and Appendix C). For the non-interacting case the von Neumann entropy is zero, and the system is therefore separable. For the solution in the impenetrable delta-function barrier limit it is equal to $S=1.056$ which is close to the maximum, $\ln(3)=1.099$, i.e., it is  close to a maximally mixed state. In Fig.~\ref{fig:Fig5} (e) and (f)  we show the OBDM for both cases, which will help interpret what a mixed state means in this system.  The diagonal of the OBDM (when $x=x')$ is the probability of finding a particle at position $x$. For the non-interacting case, it is Gaussian, as it corresponds to a single particle in a one-dimensional parabolic trap. When the interactions increase, this diagonal changes (see~\cite{garcia-march_distinguishability_2014}) and the states start to get mixed. This reflects the fact that the particles interact with each other. For the impenetrable delta-function barrier limit, two particles cannot occupy the same position along $x$ and if one is found at the center of the trap, the other two have to be slightly displaced to the edges. This explains the three-peak shape of the diagonal of the OBDM, while the appearance of structure in the off-diagonal terms indicates the presence of correlations. The classical interpretation of this is that, to determine the position of one particle,  information  about the position of the other particles is necessary, contrary to the non-interacting case. This is the essence of classical non-separability in this system.

\section{Concluding remarks}
\label{sec:conclusions}

We have shown that a quantum system consisting of three interacting atoms in one dimension with arbitrary statistics can be simulated in an optical setup. For this we have introduced a new kind of optical fiber with a GRIN refractive index profile and three thin slabs of a metallic material. Using discrete group theory we have classified the optical modes in such a fiber with appropriate modal numbers, and obtained exact solutions for the case in which the slabs are infinitely narrow and high. In the analogy with the interacting atom system the modal numbers turn into quantum numbers and, in particular, the modal number of the orbital angular pseudo-momentum together with the reflection parity play the role of the  parameters quantifying the particle statistics. We have shown that the spatial profile of the input beam permits to select the statistics of the atoms emulated in the fiber (e.g. three fermions, three fermions or mixtures).

We remark that lesser symmetries can appear in the system for specific choices of the coupling. For example, if in the BBX case the coupling constants are $g_{\rm{13}}=g_{\rm{23}}\neq g_{\rm{12}}$ the symmetry is no longer $\mathcal{C}_{6v}$ but $\mathcal{C}_{2v}$. A similar situation appears for FFX if $g_{\rm{13}}=g_{\rm{23}}\neq 0$. In this work we have restricted our study to the most general $\mathcal{C}_{6v}$ system, however the two examples of $\mathcal{C}_{2v}$ symmetric systems can also be easily accessed with a similar fiber set-up.

We have also discussed the appearance of classical non-separability in the system in the limit where the slabs are infinitely narrow and high, and where the optical states are close to a maximally mixed state. Due to the correspondence to multi-particle entanglement, this represents classical intersystem entanglement ~\cite{1998SpreeuwFoP,spreeuw_classical_2001,harshman_tensor_2007,aiello_quantumlike_2015}. It is interesting to note that one can also explore nonlocality in the setup we present by using the Wigner representation of the states \cite{Banaszek1998,Fogarty_2011,Li2011}.

The fundamental analogy between optical and quantum systems opens the door to explore more technical analogies as well. For example, there are proposals for implementations of quantum computing algorithms in optical systems~\cite{manko_quantum_2001,2002BhattacharyaPRL,2015PerezGarciaPLA}, optical implementations of teleportation protocols~\cite{2016Guzman-SilvaLPR}, and applications of the presence of classical non-separability for metrology~\cite{2014ToppelNJP,2015Berg-JohansenOptica}.   On top of this other properties defined only for the quantum system have been found in the classical ones, such as analogies to quantum (wave) chaos~\cite{doya_speckle_2002}, quantum walks~\cite{2003KnightPRA}, or classical non-separability in vector vortex beams~\cite{2016DAmbrosioPRA}. We expect that the system introduced here allows for the exploration of this kind of applications, especially when combined with polarization.

\section*{Funding}
 Spanish Ministry MINECO (National Plan15 Grant:  FISICATEAMO No.  FIS2016-79508-P, FPI); European  Social  Fund;   Fundaci\'o  Cellex;   Generalitat de  Catalunya  (AGAUR  Grant  No.2017,  SGR  1341, and CERCA Program); European Commission (ERC AdG OSYRIS, EU FET-PRO  QUIC);  National  Science  Centre,  Poland-Symfonia (Grant No.  2016/20/W/ST4/00314); TF acknowledges support under JSPS KAKENHI-18K13507, TB and TF acknowledge support from the Okinawa Institute of Science and Technology Graduate University; A.F. acknowledges the Spanish MINECO Project No. TEC2017-86102-C2-1) and Generalitat Valenciana (Prometeo/2018/098); MAGM acknowledges funding from the Spanish Ministry of Education and Vocational Training (MEFP) through the Beatriz Galindo program 2018 (BEAGAL18/00203).

\section*{Acknowledgments}
We thank P. Grzybowski and E. Pisanty for encouraging and motivating discussions.

\appendix

\section{Classification of mode symmetries}
\label{sec:sym}

For any form of two-body interaction, the three-particle Hamiltonian with harmonic trapping given in eq.~(\ref{model31}) is symmetric under the finite group of transformations given by the particle permutation symmetry of three identical (but not necessarily indistinguishable) particles combined with parity inversion about the minimum of the harmonic trapping potential. When these symmetries are restricted to the relative configuration space, they realize the point group $C_{6v}$, i.e.\ the rotation and reflection symmetries of a hexagon~\cite{harshman_symmetries_2012,garcia-march_distinguishability_2014,harshman_one-dimensional_2016}. The twelve elements of $C_{6v}$ and their realizations as transformations of relative configuration space are summarized in Table~\ref{tab:elements}. An optical fiber simulating the three-particle model will have this hexagonal symmetry in the transverse profile of the fiber. 

Subgroups of $C_{6v}$ are useful when the particles are partially distinguishable. We call attention to three subgroups in particular:
\begin{itemize}
\item The group generated by reflections across the three lines $x=0$, $x + \sqrt{3}y=0$, and $x - \sqrt{3}y=0$ is a subgroup of $C_{6v}$ isomorphic to $C_{3v}$, the symmetry of an equilateral triangle. This group has six elements and is the realization in the fiber of the permutation symmetry of three identical particles. Each reflection corresponds to a pairwise particle exchange. For example, the reflection across the line $x=0$ in the fiber corresponds to exchanging particles $1$ and $2$ in the particle model. The product of two different reflections is a rotation by $2\pi/3$ correspond to cyclic three-particle exchanges in the model.
\item The subgroup containing only the rotations in $C_{6v}$ is called $C_{6}$. This group is useful for the analysis of vortex states of light because the generator $\hat{c}_6$ of  $C_{6}$ is a rotation by $\pi/3$ and a state with OAPM $m$ transforms like $\exp(i m\pi/3)$ under this rotation.
\item There are several subgroups of $C_{6v}$ that are isomorphic to $C_{2v}$, i.e.\ the point symmetries of a rectangle. In particular, we focus on the instance of $C_{2v}$ that aligns with the $\{x,y\}$ Jacobi coordinates and includes the reflection $\hat{\sigma}_{213}$ corresponding to the exchange of particle $1$ and $2$ with eigenvalue $r=\pm 1$. The other three elements of this $C_{2v}$ subgroup are parity inversion $\hat{\pi}$, the product $\hat{\sigma}_{213}\hat{\pi}$ and the identity. This $C_{2v}$ subgroup is useful when considering the case of partially distinguishable particles like two bosons in the same spin state and one distinguishable by a different spin state. This subgroup is also relevant in more generalized models in which one of the two-body interactions is different from the other two and is evident in Fig.~\ref{fig:fig4}.
\end{itemize}

\begin{table}[ht]
\centering
\begin{tabular}{|c|c|c|}
\hline
$g\in C_{6v}$ & $g\in S_3 \times Z_2$ & $\varphi\rightarrow \varphi'$ \\
\hline
$E$ & $\hat{e}$  & $\varphi$ \\
$\sigma_{v}$ & $\hat{\sigma}_{213}$ & $-\varphi + \pi$\\
$\sigma_{v'} $ & $\hat{\sigma}_{132} $ & $-\varphi + \frac{\pi}{3}$ \\
$\sigma_{v''} $ & $\hat{\sigma}_{321} $ & $-\varphi - \frac{\pi}{3}$ \\
$C_3^{-1}$ & $\hat{\sigma}_{231}$ & $\varphi - \frac{2\pi}{3}$ \\
$C_3$ & $\hat{\sigma}_{312}$ & $\varphi + \frac{2\pi}{3}$ \\
$C_2$ & $\hat{\pi}$  & $\varphi+\pi$ \\
$\sigma_{d}$ & $\hat{\pi}\hat{\sigma}_{213}$ & $-\varphi$\\
$\sigma_{d'} $ & $\hat{\pi}\hat{\sigma}_{132} $ & $-\varphi - \frac{2\pi}{3}$ \\
$\sigma_{d''} $ & $\hat{\pi}\hat{\sigma}_{321} $ & $-\varphi + \frac{2\pi}{3}$ \\
$C_6$ & $\hat{\pi}\hat{\sigma}_{231}$ & $\varphi + \frac{\pi}{3}$ \\
$C_6^{-1}$ & $\hat{\pi}\hat{\sigma}_{312}$ & $\varphi - \frac{\pi}{3}$ \\
\hline
\end{tabular}

\caption{The first column is the symmetry transformation designated by the cor\-re\-spond\-ing element of the point symmetry group of the regular hexagon permutation group $C_{6v}$.  The second column is the same transformation expressed as the corresponding element of $S_3 \times Z_2$.  We use the notation for $S_3$ permutation group elements such that $\hat{\sigma}_p$ for $p\in S_3$. For example, $\hat{\sigma}_{213}$ exchanges particles $1$ and $2$, $\hat{\sigma}_{312}$ cycles the particles $123$ to $312$ and $\hat{\sigma}_{123} = \hat{e}$ the identity. The element $\hat{\pi}$ is parity inversion.  The third column is the equivalent transformation of the cylindrical Jacobi coordinate $\tan\varphi=y/x$.}
\label{tab:elements}
\end{table}

Because $C_{6v}$ is a symmetry of the fiber and the relative interacting Hamiltonian, energy levels are associated to its irreducible representations (irreps), whose properties are summarized in Table~\ref{tab:irreps}. There are four one-dimensional (or singlet) irreps denoted $A_1$, $A_2$, $B_1$ and $B_2$ and two two-dimensional (or doublet) irreps denoted $E_1$ and $E_2$. This means that unless some other symmetry is present, there will only be singly-degenerate or doubly-degenerate energy levels, as is demonstrated by our numerical solutions, see Fig.~\ref{fig:fig2}. Note that half of the irreps correspond to even parity states and half to odd parity states. We plot in Fig.~\ref{fig:fig3} two examples of XYZ solutions, these are very important {\it vortex}-like solutions. 

\begin{table*}[ht]
\centering
\begin{tabular}{|c|c|c|c|c|c|}
\hline
$C_{6v}$ & $[C_{3v}]^\pi$ & OAPM & $r$  & $C_{2v}$ & Possibilities\\
\hline
$A_1$ & $[3]^+$ & $m=0$ & $r=1$ & $A_1$ & BBB, BBX, XYZ \\
$A_2$ & $[1^3]^+$ & $m=0$ & $r=-1$ & $A_2$ & FFF, FFX, XYZ \\
$B_1$ & $[1^3]^-$ & $m=3$ & $r=-1$ & $B_1$ & FFF, FFX, XYZ\\
$B_2$ & $[3]^-$ & $m=3$ & $r=1$ & $B_2$ & BBB, BBX, XYZ  \\ \hline
\multirow{2}{*}{$E_1$} & \multirow{2}{*}{$[21]^-$} & \multirow{2}{*}{$m=|1|$} & $r=-1$ & $B_1$ & FFX, XYZ \\
&& & $r=1$ & $B_2$ & BBX, XYZ \\ \hline
\multirow{2}{*}{$E_2$} & \multirow{2}{*}{$[21]^+$} & \multirow{2}{*}{$m=|2|$} & $r=1$ & $A_1$ & BBX, XYZ \\
&&& $r=-1$ & $A_2$ & FFX, XYZ \\
\hline
\end{tabular}
\caption{The first column lists the irreps of $C_{6v}$ using the notation of \cite{hamermesh_group_1989}. The second column shows how irreps of $C_{6v}$ can also be described as irreps of $C_{3v}$ and parity. The group $C_{3v}$ is isomorphic to the symmetric group $S_3$ and has a totally symmetric irrep denoted $[3]$, a totally antisymmetric irrep denoted $[1^3]$ and a mixed symmetry irrep denoted $[21]$. The third column lists the OAPM that characterize the irreps of $C_6$. The fourth column give the irreps of $C_{2v}$; for the two-dimensional $C_{6v}$ irreps $E_1$ and $E_2$ there are two irreps of $C_6$ and $C_{2v}$ that appear, and they can be though of as different ways of diagonalizing the doublet. The final column gives the possible particle content of each state. BBB (FFF) means three indistinguishable bosons (fermions); BBX (FFX) two indistinguishable bosons (fermions) and one other identical but distinguishable particle; XYZ three identical but distinguishable particles.}
\label{tab:irreps}
\end{table*}

Distinguishable identical particles do not necessarily have any specific particle exchange symmetry, so they can populate any type of irrep. Indistinguishable bosons must be symmetric under pairwise exchanges and can only populate energy levels that carry the singlet irreps $A_1$ (positive parity) and $B_2$ (negative parity). Indistinguishable fermions must be antisymmetric, and can only populate $A_2$ (positive parity) and $B_2$ (negative parity) energy levels. Partially distinguishable bosons and fermions are more complicated. For example, two indistinguishable bosons and a third particle must be symmetric under the exchange of two of the particles, say particles 1 and 2, but can have any symmetry relation with the third. As a result, they effectively have $C_{2v}$ symmetry and should be in a bosonic irrep of that subgroup. See Table~\ref{tab:irreps} for a cataloging of these results.

The $C_{6v}$ symmetry is independent of the strength and exact form of the two-body interaction, and this has consequences for adiabatic or diabatic changes in the Hamiltonian. When the Hamiltonian changes, there can at most be mixing between energy levels carrying the same irrep. Therefore, the symmetry of the input beam determines which irreps, and therefore, the effective particle content of the interacting model being simulated.

\section{Bose-Fermi mapping}
\label{sec:Bose-fermi}

The exact solutions  for three bosons in a harmonic trap in the infinite delta-barrier limit (\ref{eq:refractiveindexC6-delta}) can also be derived from the Bose-Fermi mapping~\cite{girardeau_relationship_1960}, which we describe here briefly.

The non-interacting Hamiltonian (8) is a three-dimensional isotropic harmonic oscillator and can be exactly solved in many different coordinate systems. Perhaps the most obvious is the product single-particle wave functions
\begin{equation}\label{unsymmetrized_harmonic}
\phi_{n_1,n_2,n_3}(x_1,x_2,x_3) = \varphi_{n_1}(x_1)\varphi_{n_2}(x_2)\varphi_{n_3}(x_3)
\end{equation}
where $\varphi_n(x)$ is the one-dimensional harmonic oscillator energy eigenstate
\begin{equation}
 \varphi_n(x)=\left(\pi^{1/4}\sqrt{2^n n!}\right)^{-1}e^{-x^2/2}H_n(x)
\end{equation}
 with energy $\hbar\omega(n + 1/2)$ ($H_n(x)$ is the $n$th Hermite polynomial). Therefore, the state (\ref{unsymmetrized_harmonic}) has total energy $E = \hbar\omega(n_1 + n_2 + n_3 + 3/2)$.

For distinguishable particles, the quantum numbers $n_i$ can take any non-negative integer value, but for identical (or partially identical) sets of fermions or bosons, then there are restrictions on the sets of allowed $n_i$ and the states (\ref{unsymmetrized_harmonic}) must be symmetrized and antisymmetrized appropriately. The ground state for three identical non-interacting bosons is the state $\{n_1,n_2,n_3\} = \{0,0,0\}$ and remains separable, but the first excited bosonic state is the symmetric combination of the three permutations of $\{n_1,n_2,n_3\} = \{0,0,1\}$ and is not separable. 

Similarly, the ground state of fermions is the antisymmetrized superposition of six permutations of the state (\ref{unsymmetrized_harmonic}) with $\{n_1,n_2,n_3\} = \{0,1,2\}$, also known as the Slater determinant. In the ground state each fermion occupies one of the three lowest energy single-particle eigenstates, and thus the energy of the state is $E=\hbar\omega(1/2+3/2+5/2)=9/2\hbar\omega$. A bit of algebra brings the antisymmetrized expression for the fermionic ground state into a Jastrow form
\begin{equation}
\label{eq:Jastrow}
 \phi_{\rm{F,gs}}(x_1,x_2,x_3)=C\left[\prod_{i=1}^3 e^{-x_i^2/2} \right]\prod_{1\le j < k \le 3}(x_k-x_j), 
\end{equation}
 with 
 \begin{equation}
  C=2^{3/2}\left(\frac{1}{a}\right)^{3/2}\left[3!\prod_{n=0}^{2}n!\sqrt{\pi}\right]^{-1/2}. 
 \end{equation}
Because of the the factor $(x_k-x_j)$ for each pair of particles in (\ref{eq:Jastrow}), the function $\phi_{\rm{F,gs}}(x_1,x_2,x_3)$ vanishes wherever two particles coincide and changes sign as one moves across this boundary, as one expects for antisymmetrized fermionic states (see Fig.~\ref{fig:fig3}). Excited fermion states $\phi_{\rm{F,exc}}(x_1,x_2,x_3)$ can be constructed either by using Slater determinants of states with set of three distinct quantum numbers higher in energy that $\{0,1,2\}$ or, by analogy with (\ref{eq:Jastrow}), finding higher-order totally antisymmetric polynomials of three variables.

Fermionic states have nodes whenever $x_i= x_j$, so they do not ``feel'' the $\delta$-function two-body interaction and therefore non-interacting fermionic energy eigenstates are also energy eigenstates of the Hamiltonian (9) in the limit when the interactions are zero-range. Some algebra demonstrates that when restricted to the relative coordinate, the ground state  wave function (\ref{eq:Jastrow}) has the same form as the eigenmode $|m=3, \tilde{\nu} =0 ,\tilde{\ell} =0 \rangle$ and the first excited state constructed from the Slater determinant of  $\{0,1,3\}$ corresponds to $|m=0, \tilde{\nu} =0 ,\tilde{\ell} =1 \rangle$. After that the identification gets more complicated because of degeneracies.

The Bose-Fermi mapping theorem allows one to construct the exact solution for three interacting identical bosons with delta-barrier interactions and $g\to\infty$ from the exact solution for non-interacting fermions~\cite{girardeau_relationship_1960}. This is easily fulfilled for the ground state taking the modulus of the solution,  $\phi_{\rm{B,gs}}=|\phi_{\rm{F,gs}}|$, which gives rise to 
\begin{equation}
\label{eq:Jastrow2}
 \phi_{\rm{B,gs}}=C\left[\prod_{i=1}^3 e^{-x_i^2/2} \right]\prod_{1\le j < k \le 3}|x_k-x_j|. 
\end{equation}
The excited bosonic states are obtained in the same way from those of the excited fermions by defining a symmetrization function $A(x_1,x_2,x_3)=\prod_{j>i}\mbox{sign}(x_j-x_i)$, with sign$(x)$ the sign function to adjust the relative phases. Then any excitation is obtained as $\phi_{\rm{B,exc}}=A \phi_{\rm{F,exc}}$. In the relative eigenmode basis $|m, \tilde{\nu} ,\tilde{\ell} \rangle$, this mapping is equivalent to exchaning the OAPM labels $m=0$ and $m=3$ on the FFF and BBB states.

\end{document}